# Linking the pressure dependence of the structure and thermal stability to $\alpha$- and $\beta$-relaxations in metallic glasses

**Jie Shen[1,2†*], Antoine Cornet[1,2†], Alberto Ronca[1,2], Eloi Pineda[3], Fan Yang[4], Jean-Luc Garden[1], Gael Moiroux[1], Gavin Vaughan[2], Marco di Michiel[2], Gaston Garbarino[2], Fabian Westermeier[5], Celine Goujon[1], Murielle Legendre[1], Jiliang Liu[2], Daniele Cangialosi[6,7] and Beatrice Ruta[1,2*]**

[1] *Institut Néel, Université Grenoble Alpes and Centre National de la Recherche Scientifique, 25 rue des Martyrs - BP 166, 38042, Grenoble cedex 9, France*

[2] *European Synchrotron Radiation Facility, 71 avenue des Martyrs, CS 40220, Grenoble 38043, France*

[3] *Department of physics, Institute of Energy Technologies, Center for Research in Multiscale Science and Engineering, Universitat Politècnica de Catalunya - BarcelonaTech, 08019 Barcelona, Spain*

[4] *Institut für Materialphysik im Weltraum, Deutsches Zentrum für Luft- und Raumfahrt (DLR), 51170 Köln, Germany*

[5] *Deutsches Elektronen-Synchrotron DESY, Notkestraße 85, D-22607 Hamburg, Germany*

[6] *Donostia International Physics Center, Paseo Manuel de Lardizabal 4, 20018 San Sebastián, Spain*

[7] *Centro de Física de Materiales (CSIC–UPV/EHU), Paseo Manuel de Lardizabal 5, 20018 San Sebastián, Spain*

[†] These authors equally contributed

*corresponding authors

jie.shen@neel.cnrs.fr

beatrice.ruta@neel.cnrs.fr

## ABSTRACT

Glasses derive their functional properties from complex relaxation dynamics that remain enigmatic under extreme conditions. While the temperature dependence of these relaxation processes is well-established, their behavior under high-pressure conditions remains poorly understood due to significant experimental difficulties. In this study, we employ cutting-edge experimental techniques to probe the pressure evolution of the relaxation spectrum in a $Zr_{46.8}Ti_{8.2}Cu_{7.5}Ni_{10}Be_{27.5}$ metallic glass across gigapascal pressure ranges. Our findings reveal two distinct relaxation mechanisms under high pressure: In the $\beta$-relaxation regime, compression drives the system with reduced atomic mobility and enhanced structural disorder, without significant density changes. Conversely, $\alpha$-relaxation under pressure promotes density-driven structural ordering that improves thermal stability. Notably, the transition between these regimes occurs at a constant $T/T_{g,P}$ ratio, independent of applied pressure. These results provide crucial insights for decoupling the competing structural and relaxation contributions to glass stability, establishing a systematic framework for tailoring glass properties through controlled thermo-mechanical processing.

## TEASER

Hydrostatic pressure creates unique metallic glass states by independently controlling $\beta$- and $\alpha$-relaxations.

## INTRODUCTION

Understanding the relaxation spectrum of glass-formers is a great challenge in condensed matter physics and materials science with important technological implications, as relaxation modes play a prominent role in the aging and performance of glasses (1-5). The structural relaxation process, also called $\alpha$-relaxation, is responsible for the viscous flow in the liquid phase and it involves large cooperative rearrangements (2,6). Its temperature evolution in supercooled liquids determines the kinetic fragility of the material, a quantity that correlates with a large variety of material properties and allows classifying glass-formers in fragile and strong systems (7). The secondary $\beta$-relaxation is instead related to the thermal activation of localized atomic mobility and persists also at temperatures lower than the glass transition temperature, $T_g$ (8-14). Additional secondary processes, such as $\gamma$- and $\beta'$-relaxations, have been also identified in several glasses (15-17), and have been attributed to the diffusion of liquid-like atoms far below $T_g$ (18,19). Numerous studies have focused on the strong dependence of the macroscopic properties and structure of glasses on various relaxation dynamics, which are governed by changes in temperature (4,20-27).

Hydrostatic pressure, as an important external factor, also play a unique role in the relaxation process. In molecular, ionic, and polymeric systems, dielectric spectroscopy experiments indicate that the $\alpha$-relaxation considerably slows down under high pressure, shifting $T_g$ to larger values (28-30). In contrast, secondary $\beta$-relaxation processes are less sensitive to pressure and exhibit a different response to isothermal compressions and isobaric thermal treatments. This different response leads to an increasingly larger difference in time scales between $\alpha$- and $\beta$-relaxation on increasing pressures, and to the impossibility to rescale the whole dynamics in a single master curve (31,32).

Due to the difficulty to perform experiments at *in-situ* high-pressure conditions, the pressure dependence of relaxation processes in metallic glasses (MGs) remains poorly understood. Furthermore, because of their highly packed structure, MGs are less sensitive to pressure compared to molecular or network glasses (33), requiring the application of large pressure values—in the GPa range—to observe noticeable changes in their properties (34-37).

To circumvent the experimental challenges of performing *in-situ* high-pressure measurements, many studies have focused on the pressure response of pre-compressed samples. In these cases, the material is first compressed at a given pressure and temperature, and subsequently studied after recovering the resulting glass at ambient conditions. As the recovered materials contain information on the glass ability to retain permanent changes induced by compression after decompression, this method presents also the advantage to produce new amorphous materials of the same composition but with different properties, which can be tuned for the desired applications. For example, $SiO_2$ glass exhibits a density increase of over 20% at ambient conditions when previously subjected to pressures greater than 21 GPa (33).

In the case of MGs, a 0.6-2% increase in density has been reported in samples previously compressed at room temperature at 5.5-20 GPa (33,38,39), while recent *in-situ* high pressure X-ray diffraction (XRD) experiments have indicated the absence of important permanent structural changes in Zr-based MG even when compressed up to 30 GPa at room temperature (40). From the thermodynamic point of view, studies have shown that isobaric quenching from the supercooled liquid, or prolonged annealing near $T_g$ under high pressure, can significantly enhance the thermodynamic stability of MGs; additionally, $T_g$ increases with pressure at a rate of $dT_g/dP$=3.6–10 K/GPa (36,41,42). In contrast, most MGs compressed in the glassy state are typically found to be in a higher energy state than the pristine samples (38,39,43-46).

Likewise crystalline materials, in some compositions pressure can even modify the structure of glass-formers and drive them through a transition toward a different amorphous state. Such polyamorphic transitions from a low to a high-density amorphous state have been reported mainly in covalent and hydrogen-bonded systems, such as chalcogenide glasses (47), amorphous ice (48) and $SiO_2$ (49), where it has been associated to changes in the coordination number leading to a high randomly packed structure above the transition. Polyamorphism under high pressure has been identified also in MGs. In Ce-based MGs this glass-glass transition has been associated to an electronic transition in the *f*-shell which is enhanced by shell size reduction at high pressure (50-52). Differently, in Zr-based glass formers, Dmowski *et al.* (53) have shown the possibility to obtain MGs with a highly-packed random structure similarly to covalently bonded glass formers, by quenching at high pressure a high-density phase occurring in the supercooled liquid.

All these works show glass formers are influenced not only by the microscopic ongoing relaxation processes but also by structural mechanisms under hydrostatic pressure. Still, to the best of our knowledge, these different contributions have not been extensively investigated, although this being fundamental to provide a comprehensive picture of the pressure effect on glass-formers. In this study, we conducted a systematic study of the thermal response and structure of a series of $Zr_{46.8}Ti_{8.2}Cu_{7.5}Ni_{10}Be_{27.5}$ (Vit4) MGs previously compressed at various pressures in the 1-7 GPa range, and for temperatures covering cold compression at ambient temperature, hot compression in the glassy state, and compression in the supercooled liquid phase. By analyzing the activation energies, our approach allows us to identify the pressure response of both the $\alpha$- and $\beta$- relaxations and to clarify their effect on the thermal stability and structure of the recovered glasses. We find that activation energy data measured at different pressures can be rescaled onto a single master curve when normalized by $T/T_{g,P}$. This approach falls short for the evolution of the thermodynamic state of the different compressed glasses which instead is influenced also by specific structural rearrangements occurring during the compression. Our study highlights the critical response to hydrostatic pressure of a prototypical MG, providing clear thermo-mechanical processing guidelines for designing glasses with the same composition but different properties.

## RESULTS

### Pressure dependence of the relaxation spectrum

In compression experiments, the properties of the material are influenced by both the applied pressure, $P_{comp}$, and compression temperature, $T_{comp}$. To explore the effect of $T_{comp}$, we first employed an isobaric densification protocol and prepared different samples by maintaining a constant $P_{comp}$ of 7 GPa and varying $T_{comp}$ from 298 to 693 K (from the deep glass state to the fully-equilibrated supercooled liquid phase, **Fig. 1A**). This process allows obtaining different glasses, which were subsequently studied by single shot flash differential scanning calorimetry (FDSC) and synchrotron XRD. **Fig. 1B** and **1C** display FDSC curves of the compressed glasses. In a single shot FDSC experiment, the sample is measured without the usual melting on chip and is only heated into the supercooled liquid region in order to reveal the kinetic path the thermal changes induced by densification to recover the supercooled liquid. The validity of this approach as well as the reproducibility of the data is discussed in the methods and supplementary materials (see also **Fig. S1**). The scans were performed using a heating rate, $\Phi_h$, of 200 K/s, much faster than the cooling rate of 20 K/min employed during the densification protocols. In this way it is possible to generate an amplified overshoot peak in the curves (54), enabling a better comparison of the thermodynamic state among the different samples. To highlight the effect of the heating/cooling rates on the scans, in **Fig. 1B** and **1C**, we show also the heating scan of a sample obtained from the supercooled liquid with a quenching rate of 200 K/s, and immediately heated at the same rate (black lines). These scans markedly differ from all the other curves, which have been instead obtained using a cooling rate much lower than the heating rate employed during the measurements.

To investigate the effect of pressure on the thermal response and disentangle pressure effects from temperature dependencies of the measured observables, we first compare the compressed glasses to systems subjected to the same thermal treatments but at the ambient pressure of 1 atm. With respect to the glasses pre-treated at 1 atm, $T_{g,onset}$ shifts by 6 K to larger values in the samples compressed at 7 GPa from $T_{comp} \geqslant$ 643 K (**Fig. 1B**). We remind that $T_{g,onset}$ is here defined as the intersection point between the glass baseline and the tangent drawn at the midpoint of the glass transition step. The increase in $T_{g,onset}$ suggests an increased glass stability. Moreover, the independence of the calorimetric response from the choice of $T_{comp}$ (≥643 K) suggests the sample is at equilibrium during the densification protocol. As shown in **Fig. 1B**, all curves overlap on top of each other as it occurs also for the reference glasses which are all quenched from the corresponding supercooled liquid at 1 atm ($T_{g,1\ atm}$=596 K, (55)).

In contrast, when the annealing temperature, $T_a$, or $T_{comp}$ is below the corresponding $T_g$ at each pressure, the thermodynamic state of the resulting glasses varies with temperature, reflecting the response of different metastable glassy states formed during sample preparation (**Fig. 1C**). At 1 atm, lowering $T_a$ shifts the main endothermic peak to a sub-$T_g$ endothermic contribution, corresponding to the curve change from $T_a$=583 K to 513 K in **Fig. 1C**. This behavior is common in the aging process of MGs (13,56,57), inorganic glasses (58), and polymeric glasses (59,60). It is attributed to more localized particle motion driven by $\beta$-relaxation in the deep glassy state, and is also referred to as the shadow glass transition (58,61). In the glasses compressed at 7 GPa at $T_{comp}$=583 K and 513 K, the sub-$T_g$ endothermic peak appears with significantly reduced intensity with respect to the corresponding data measured at 1 atm. In the extreme case of $T_{comp}$=298 K (well below $T_g$), the heat flow curve is almost insensitive to pressure and shows a slightly broader exothermic peak compared to the as-cast sample.

By using Moynihan's area-matching method in the calorimetric scans (62), we can confirm the previous observations from the evaluation of the fictive temperature, $T_f$, which can be considered as a metric of the thermal stability of the glass (13,24,63). It is worth noting that $T_f$ in this work is used as a parameter to describe the thermodynamic state of decompressed samples (see **Fig. S3** for further details) and does not reflect the real temperature at which vitrification occurs under *in-situ* high pressure. **Fig. 2A** reports the $T_f$ obtained in glasses compressed at 7 GPa and different temperatures between 298 K and 693 K. Reference values for glasses annealed at the same temperatures and 1 atm are reported as well. In both sets of glasses, $T_f$ evolves with the annealing or compression temperature from a maximum value at low temperatures, signature of a highly energetic frozen glass configuration of the as-cast samples, to an increasingly lower value at larger $T_a$ and $T_{comp}$, a consequence of thermally-activated microscopic structural rearrangements occurring in the material (63). It is immediately evident that the choice of the temperature, $T_{comp}$, employed during the compression can lead both to larger or lower values of $T_f$ with respect to the reference samples annealed at 1 atm. For $T_{comp}$ lower than 400 K, we observe a constant increase in $T_f$ with pressure independently on the applied temperature during the densification, suggesting the occurrence of a weak rejuvenation at 7 GPa with respect to the 1atm samples. In glasses compressed between 400 K and 630 K, $T_f$ are systematically larger by ~30 K (compared to samples annealed at the same temperature under 1 atm). In contrast, for $T_{comp}\geq$643 K $T_f$ decreases to a $T$-independent lower value (by 3-4 K) with respect to the 1 atm data. The lower $T_f$ value is further evidenced by the increased area of the endothermic peak (**Fig. S4**). The glass obtained by high pressure quenching from the supercooled liquid phase, not only show a higher $T_{g,onset}$ (**Fig. 1B**), but also a lower $T_f$, reflecting a more stable state, which aligns with the larger density reported in high-pressure quenched MGs (53,64).

To get the activation energies associated with the different relaxation processes, **Fig. 2B** shows Kissinger plots of glasses annealed and compressed under different $T_a$ or $T_{comp}$ and then measured with FDSC at different heating rates (56,65). In this figure, each data point corresponds to a new sample. The corresponding activation energies obtained by fitting the data are shown in **Fig. 2C**. The activation energies distinctly evolve between two extreme values on increasing $T_a$ in the reference glasses or $T_{comp}$ in the compressed MGs. These values are close to those reported in literature for the $\beta$- and $\alpha$-relaxation processes, respectively (~28 $RT_{g,1atm}$ for the $\beta$- relaxation, and ~42 $RT_{g,1atm}$ for the $\alpha$-relaxation (9,56,57,66)). As shown in **Fig. 2C**, pressure shifts the crossover between the two values by ~30 K at 7 GPa. This means that applying hydrostatic pressure at temperatures slightly below $T_{g,1atm}$ can induce a transition in the dominant relaxation mode, explaining thus the difference between the annealing at 1 atm and at 7 GPa. At 583 K, for instance, the significant decrease in activation energy for the samples compressed at 7 GPa suggests that the $\alpha$-relaxation is notably suppressed, leaving only the unrestricted localized atoms active, contributing to the $\beta$-relaxation observed as the sub-$T_g$ endothermic peak. At the same temperature, however, the dynamics at 1 atm is still governed by the $\alpha$-relaxation, which is manifested as a post-$T_g$ endothermic peak. For $T_{comp} \geq$ 643 K, the constant activation energy indicates that $\alpha$-relaxation is fully activated also at 7 GPa. The $\alpha$-relaxation activation energy for the sample compressed from the supercooled liquid phase is about 4% lower compared to that of the samples quenched at 1atm. We attribute this effect to a pressure-induced decrease in the fragility of the corresponding supercooled liquid (67,68). The lower $E_a$ of samples densified by quenching from the supercooled liquid at 7 GPa indicates that the corresponding supercooled liquid becomes "stronger" with pressure, as expected considering the reduced atomic mobility. This interpretation is further supported by the broadening of the glass transition overshoot peak observed in the FDSC scans (**Fig. S4**), indicating a wider glass transition interval in the 7 GPa glass. This is another hallmark of a stronger supercooled liquid, where the transition from the glass to the supercooled liquid occurs gradually (69).

Comparing the $T_f$ and $E_a$ values, we can divide the $T_{comp}$ space into three regions: (I) the cold compression region where temperature effects are almost negligible in this composition. Here, the slightly larger exothermic peak area after compressions indicates a purely mechanical rejuvenation; (II) the temperature-pressure coupling region: here, as discussed in detail later, the difference in $T_f$ between the compressed and 1 atm samples is amplified with the activation of $\beta$-relaxation, but decreases with the activation of $\alpha$-relaxation; (III) the equilibrium region, accompanied by a complete activation of the $\alpha$-relaxation: here compression leads to relaxation and is manifested by a decrease in $T_f$. In this region, $T_f$ and $E_a$ are constant, indicating that the system has reached a state of thermodynamic equilibrium for $T \gtrsim 625$ K. This means that during the densification process the transition between the glass and the supercooled liquid occurs between 625 K and 643 K (where equilibrium is achieved). Taking the midpoint of 634 K, we designate it as the $T_{g,P}$ under *in-situ* compression at 7 GPa, which would correspond to an increase of ~38 K with respect to the value at 1 atm. This increase is the consequence of the slowing down of the $\alpha$-relaxation process with pressure. Given that the activation energy of the $\alpha$-relaxation decreases by only ~4% under high pressure, it remains reasonable to assume a similar temperature dependence of the corresponding structural relaxation time, $\tau_\alpha$, at both pressures. Considering thus just a temperature shift of the 1 atm Vogel-Fulcher-Tamman function used to describe the temperature dependence of $\tau_\alpha$ (55), the increase in $T_g$ would correspond to a slowdown in the dynamics by a factor of ~1500 at 7 GPa (see **Fig. S5**). This estimation is orders of magnitude smaller than the values reported in molecular glass-formers (29,30), reflecting the lower sensitivity to pressure of MGs (35-37).

We find that the pressure dependence of the $\alpha$-relaxation process is responsible for the shift to larger temperatures of the crossover between the $\alpha$- and $\beta$-dominated relaxation dynamics at 7 GPa. This is shown by the perfect scaling into a single master curve of the activation energy data at different pressures when the temperature is normalized by the corresponding $T_g$ values (**Fig. 2D**). This means that independently of the applied pressure, the activation of $\beta$- and $\alpha$-relaxations during the 10 minutes of annealing time always occurs at the same degree of equilibration, defined here as the distance from the supercooled liquid which corresponds to 0.85–0.96 $T_{g,P}$ in the studied composition.

Interestingly, while the pressure dependence of the activation energies can be scaled onto a single curve, this is not the case for $T_f$. **Fig. 3A** shows that, even when normalized by $T_{g,P}$ , there is still a large difference in the $T_f$ values obtained at the two pressures. The difference between $T_{f,\ 7GPa}$ and $T_{f,\ 1atm}$ reaches its maximum at approximately 0.85 $T_{g,P}$, while it decreases again to a minimum value once the samples enter the supercooled liquid phase. We note that the largest difference in fictive temperature corresponds to the beginning of the $\beta$-to-$\alpha$ relaxation transition. To highlight this effect, we compare FDSC curves corresponding to the same $T/T_{g,P}$ value and thus the same degree of equilibration (**Fig. 3B-3D**), while corresponding to different dynamical ranges (see also **Fig. S6**). At ~0.87 $T_{g,P}$, the $T_f$ of the 7 GPa compressed sample is significantly higher than that of the 1 atm sample. with both samples exhibiting the shadow glass transition in the FDSC curves, indicating that the $\beta$-relaxation occurred. The results show that high-pressure samples not only exhibit a significant reduction in peak intensity but also a shift of the peak to lower temperatures (**Fig. 3B**), suggesting that changes in the atomic environment under high pressure suppress the activation of $\beta$-relaxation processes in MGs. This difference gradually decreases with increasing $T/T_{g,P}$, due to the larger contribution of the $\alpha$-relaxation which is already taken into account in the comparison by the temperature normalization. At ~0.95 $T_{g,P}$, both samples show the post-$T_g$ endothermic peak instead of the shadow glass transition in the FDSC curves, and at this point, the difference in $T_f$ between the two samples decreases. In the fully $\alpha$-relaxation regime, e.g., at ~1.1 $T_{g,P}$, the endothermic peak of the high-pressure samples shifts to higher temperatures. In contrast to the $\beta$-relaxation-dominated regime, $T_f$ is slightly lower for the high-pressure samples, an effect that is visually reflected by the increased area of the endothermic peak (**Fig. S4**). The failure of the scaling for data acquired at the same degree of equilibration $T/T_{g,P}$ implies that the variation of $T_f$ with pressure cannot be solely attributed to the reduced aging caused by the increase in $T_g$. The thermal stability clearly depends also from additional contributions coming from the response to pressure of the involved relaxation processes.

**Pressure dependence of the structure**

To identify potential structural changes induced by the compression protocols, we take advantage of the high-resolution of synchrotron XRD to study the structure of compressed glasses in both the reciprocal and real space. The static structure factor reported in **Fig. 4A** shows the presence of irreversible structural rearrangements in the sample compressed at 7 GPa. To make a quantitative comparison, we calculated the center position of the first sharp diffraction peak (FSDP, **Fig. 4B**), $q_1$, and the full width at half maximum (FWHM). In MGs, smaller $q_1$ usually indicates lower density (70), while larger FWHM suggests weaker atomic correlations and faster decay of spatial coherence, often interpreted as increased structural heterogeneity (71). The corresponding data are reported in **Fig. 4D** and **4E**. Two regimes can be clearly identified. Both parameters are almost constant up to $T_{comp}$~580 K, while they shift to larger $q$ values or smaller FWHM at further increasing densification temperatures. The crossover between these two regimes correlates with the change in the microscopic process active during the compression (**Fig. 4C**) and the evolution of the fictive temperature (**Fig. 4F**). The smaller $q_1$ and larger FWHM values at low compression temperatures suggest that high-pressure annealing results in glasses with a less packed and more heterogeneous structure with respect to the glasses obtained by high-pressure quenching from the supercooled liquid phase. The crossover between the two structural regimes occurs in the $\beta$- to $\alpha$-relaxation transition region, where $q_1$ starts to increase. In this transition region both relaxation processes should coexist. Looking carefully at the data, both $T_f$, $q_1$ and the FWHM show a slightly slower evolution with increasing $T_{comp}$ between ~580 K and ~600 K (**Fig. 4D**). This subtle anomalous evolution could be the result of the competition between the two opposite physical mechanisms. Further investigations with more diffraction data under this $T_{comp}$ and pressure conditions are required for this verification. After this region, $q_1$ increases and the FWHM decreases, meaning that in this $T_{comp}$ range, pressure promotes the formation of a more packed and homogenous structure. In the supercooled liquid phase, $q_1$ and FWHM are almost independent of $T_{comp}$ within the error bars, which is consistent with the temperature invariant thermodynamic stability (**Fig. 1B** and **2A**), indicating that the sample reached an equilibrium state under compression. Interestingly, the rapid variation in $E_a$ during the activation of the $\alpha$-relaxation is reflected in both structural parameters, which shows a steeper evolution with $T_{comp}$, in agreement also with the rate of change in fictive temperature (**Fig. 4F**). These observations are consistent with molecular dynamics simulations which indicate that isothermal compression in the supercooled liquid slows down the liquid dynamics and promotes the formation of locally ordered atomic structures (35).

For completeness, **Fig. 4D** and **4E** include also structural values corresponding to glasses annealed at 1 atm and selected $T_a$. Several aspects are evident at a first glance. First, as already discussed, the compressed glasses have a different structure. Secondly, the structure of the annealed glasses is almost insensitive to the applied thermal treatment, and only a small increase in $q_1$ is observed at larger $T_a$, reflecting a more compact structure in agreement with previous studies (72,73). These results have important consequences. The different structure of the two sets of data at 1 atm and 7 GPa simply signals the occurrence of important rearrangements in the local structure under pressure. In addition, when only temperature is applied to the glass, the structure is almost insensitive, being the structure of a supercooled liquid strongly resembling that of the corresponding glass. Differently, when the heating is performed under pressure, the structural evolution of the material depends on the response of different relaxation processes to pressure: the activation of $\alpha$-relaxation drives structural densification, while $\beta$-relaxation primarily leads to local structural rearrangements and increased disorder.

To get further insights on the changes inferred by the pressure during the densification processes, we performed additional XRD measurements to get pair distribution function (PDF) analysis. The pair distribution functions, $G(r)$, obtained from the Fourier transform of the static structure factors are powerful observables to reveal the detailed atomic structure (particularly on the short-to-medium range scale, see Methods section). To achieve high resolution in real space and reliable data for the first coordination shells, the diffraction patterns were collected covering a broad *q*-range in reciprocal space up to 30 Å$^{-1}$. **Fig. 5A** shows $G(r)$ for three samples previously compressed at 7 GPa at three representative temperatures: i) $T_{comp}$= 298 K representing the cold-compression at ambient temperature, ii) $T_{comp}$= 583 K which corresponds to the onset of the *β*- to *α*-relaxation transition region, and iii) $T_{comp}$= 643 K where the system is in the supercooled liquid phase and the *α*-relaxation is fully activated. Taking the $G(r)$ of the cold-compressed sample as a reference, the differences in $G(r)$ for the samples compressed at 583 K and 643 K reveal the respective contributions of *β*- and *α*-relaxation. The inset in **Fig. 5A** presents $\Delta G(r)$, obtained by subtracting the $G(r)$ of the cold-compressed sample from that of the high-temperature compressed samples. Variations in the intensity of different coordination peaks are observed, with the most significant changes concentrated in the first three atomic shells (see also **Fig. 5B** to **5D**).

The first atomic coordination peak consists of a sub-peak at approximately 2.7 Å and a main peak at around 3.1 Å, resulting from the nearly discrete bond length distribution of nearest-neighbor atoms. The known interatomic distances indicate that the sub-peak at 2.7 Å is primarily attributed to Zr-Cu pairs, while the main peak at 3.1 Å corresponds mainly to Zr-Zr pairs (74). Compared to the cold-compressed sample, the sample densified at 583 K has a similar structure. The sub-peak remains nearly unchanged, while the main peak shows a slight increase in intensity with no shift in position. Differently, the glasses compressed at 643 K have a much lower sub-peak intensity whose position is slightly shifted to larger distances, while the main peak becomes more intense and shifts to shorter values. Based on these changes, we infer that when only *β*-relaxation is active, atomic positions undergo only minor adjustments, and the primary atomic pairs remain unchanged.

For the sample compressed at 643 K, the shortening of the radial distance is also evident at the second and third coordination peaks, and reflects an increased sample density during high-pressure quenching from the supercooled liquid, as already observed in similar Zr-based compositions (53,64). The more packed and ordered structure of the glass compressed at 643 K is confirmed also by the more pronounced oscillations of each coordination shells (with increased peak intensity), consistent with the narrowing of the FSDP (**Fig. 4D**). This shortening of the Zr-Zr bond length and the consequent readjustment of the Cu-Zr bonds under pressure have been already reported in previous experimental studies in other Zr-based MGs (53) where they have been associated to a transition from a covalent-like bond dominated structure originating from partially filled d-electron orbitals of Zr atoms that promote locally favored directional structures, to a dense random packed atomic configuration with reduced distortion in the Zr-centered coordination shells (75).

Starting from the fourth coordination shell, the diffraction signal significantly weakens, and the amplitude of $\Delta G(r)$ becomes no longer reliable for meaningful comparison (see inset of **Fig. 5A**). Therefore, to avoid potential noise interference, we refrain from analyzing the structural features at larger length scales in this study.

To clarify the differences in the structural rearrangement mechanisms occurring in the two distinct dynamical regimes described above, we investigated the pressure dependence of glasses compressed at different values of $P_{\text{comp}}$ in both the $\beta$-dominated regime at $T_{\text{comp}}$ = 583 K and in the liquid controlled by structural $\alpha$-relaxation process at 643 K. The intensity profiles of the maximum of the FSDP of the recovered glasses are shown in **Fig. 6A** and **6B** together with reference data of samples pre-annealed at 583 K and 643 K at 1 atm (black lines). Remarkably, we can clearly distinguish by eyes differences in the profile of the pre-compressed samples depending on whether the samples were compressed from the liquid or glassy state. While constant $q_1$ and FWHM are observed after compression in the glass, the $q_1$ increases and the FWHM decreases with larger pressure values in the glasses compressed from the liquid phase. These effects are quantified in **Fig. 6C** and **6D**, where we report the evolution of the cube of the $q_1$ of the FSDP, $q_1^3$ and the FWHM. For a better comprehension, we report also data obtained with XRD during *in-situ* compression at high pressure with a diamond anvil cell (see Methods section). These data provide information on the compressibility of the sample at room temperature.

The glasses recovered after compression in the liquid at $T_{comp}$=643 K, exhibit an increase in $q_1$ and a linear relationship with $\left(\frac{\partial q_1}{\partial P_{comp}}\right)_{T_{comp}} \approx 0.0008$ Å$^{-1}$/GPa within the applied $P_{comp}$ range, without showing signs of saturation in densification. This behavior mirrors the increasing densification occurring in the sample kept at high pressure, being the quantity $1/q_1^3$ often associated to density changes (72). To roughly estimate the respective proportions of elastic deformation and permanent densification during the compression process, we referred to the *in-situ* data. As shown in **Fig. 6C**, the evolution of $q_1^3$ can be fitted using a third-order Birch-Murnaghan equation of state (BM-EOS) (76), yielding an isothermal bulk modulus $B_0$=102±3 GPa, with the pressure derivative $B'$ fixed at 3.6 GPa. When $P_{in\text{-}situ}$ reaches 7 GPa, $q_1^3$ increases by approximately 6.3% compared to the pristine sample, while the 7 GPa sample compressed in the liquid shows only about 0.6% permanent densification, indicating that most of the deformation is elastic and is recovered during the pressure release process. This result is consistent with previous studies where MGs compressed under similar densification protocols exhibited density increases of 0.6%–1% (53,64). Moreover, as the pressure increases, densification is expected to continue following the linear relationship between $q_1$ and $P_{comp}$, until densification reaches saturation. Referring to previous findings, the saturation of densification in glasses typically requires pressures exceeding 20 GPa, and the maximum densification of MGs can reach up to 2% (33).

In contrast, when compressed at 583 K, the resulting samples exhibit a lower $q_1$ value, even lower than that of the 1 atm reference sample, in agreement with the data in **Fig. 4** and literature studies (42,43). This is consistently observed for different $P_{comp.}$, with $q_1$ remaining constant at an average value of 2.624±0.001 Å$^{-1}$, representing a decrease of 0.17% compared to the 1 atm reference sample. However, this observation is insufficient to confirm a decrease in density in the compressed sample. This is because the PDF analysis shows the absence of important changes in the positions of the different coordination shells in glasses compressed below their calorimetric glass transition (**Fig. 5** and **Fig. S7**). The observed structural rearrangements are instead probably the consequence of a mechanically-induced redistribution of the Zr-Zr and Cu-Zr atomic bonds at the level of the short-range order as suggested also from a comparison with a glass quenched at 1 atm (see **Fig S7**).

The lowering of $q_1$ after high pressure annealing suggests the existence of a pressure point between 1 atm and 1 GPa at which the pressure-induced structural changes saturate. However, it remains unclear whether this peak shift results from a continuous pressure-dependent evolution that eventually saturates or represents an abrupt transition to a new configuration at even lower pressures. Future synchrotron studies with different high-pressure technologies will be necessary to investigate the low-pressure range and clarify this aspect. Interestingly, the presence of distinct structural mechanisms emerges also in the behavior of the FWHM shown in **Fig. 6D**. At $T_{\mathrm{comp}}$=643 K, FWHM decreases as pressure increases, aligning with the trend observed for $q_1$, which indicates that densification and structural ordering occur simultaneously. In contrast, at $T_{\mathrm{comp}}$=583 K, the FWHM remains unchanged within the fitting error range, suggesting that in the glassy state, the structure of compressed glasses is independent of the applied pressure.

## DISCUSSION

The densification experiments discussed in this work allow us to identify the different contributions influencing the response of Vit4 glasses to hydrostatic high-pressure compressions and further elucidate their relationship with the relaxation processes.

As in other families of glass formers, pressure leads to a slowing down of the $\alpha$-relaxation process also in MGs resulting in an increase in $T_g$ of about 38 K at 7 GPa in the studied composition. The observed pressure-induced change in $T_g$ is almost one order of magnitude lower than that reported in polymeric and molecular glasses (28-32), reflecting the densely-packed nature of MGs and their consequent lower propensity to pressure-induced changes (33-37). As a direct consequence of the increase in $T_{g,P}$, the transition between the low temperature $\beta$-dominated regime and the high temperature collective motion associated to the $\alpha$-relaxation shifts to larger temperatures. Independent of the applied pressure, such crossover starts to occur at the same degree of equilibration of ~0.85 $T/T_{g,P}$, which is reflected in a visible change in all thermal and structural parameters after compression. This leads to the possibility to rescale the activation energies of the $\alpha$- and $\beta$-relaxation under different pressures into a single master curve when normalized by the corresponding $T_{g,P}$ (**Fig. 2D**). However, such scaling falls short for the fictive temperature (**Fig. 3A**). This failure indicates that the difference in $T_f$ at 7 GPa and 1 atm cannot be solely attributed to the slowing down of relaxation processes under pressure, as it is influenced also by irreversible structural reorganizations whose nature is depending on the dominant relaxation processes activated during the compression.

Cold compression at room temperature has a weak influence on the glass thermodynamic state, resulting in a tiny mechanical rejuvenation with respect to the pristine material, as signaled by the small increase in $T_f$ (**Fig. 2A** and **4F**). This effect is the consequence of the absence of long-range atomic displacements during compression and to the occurrence of structural changes in the local environment leading mainly to a redistribution of the Zr-Zr and Zr-Cu bonds at the level of the short-range order, resulting in a more disordered structure with respect to the pristine material (**Fig. 4E**).

When temperature is increased in the $\beta$-relaxation regime, thus for $T_{comp} \lesssim 0.85\ T_{g,P}$, the ratio of $T_{f,7\ GPa}/T_{f,1\ atm}$ increases (**Fig. 4F**), reaching a maximum around 0.85 $T_{g,P}$. This means that for samples annealed at the same temperature but at higher pressure, the samples are less relaxed than they would be if the pressure was absent. In our case, the high pressure reduces the free volume, and therefore atoms exhibit lower mobility. We stress that this effect is very different from rejuvenation, which, instead, would have resulted in the evolution of the system to a more unrelaxed state due to the high pressure (46). The latter scenario would be unexpected if we also consider the additional aging effect under high temperature (36). From the structural point of view, the $G(r)$ change of the high-pressure annealed glass (relative to the cold-compressed state) reveals only tiny modifications in the local environment of Zr atoms with increasing $T_{comp}$, as indicated by the enhanced intensity of the main peak of the first coordination shell (**Fig. 5B**). The absence of noticeable shifts in the positions of the coordination shells suggests that the density of the material is essentially unchanged (**Fig. S7**).

With increasing temperature, liquid-like collective atomic dynamics begins to be activated from ~0.85 $T_g$ (19,77-79). Our data show that this characteristic temperature corresponds to the onset of $\beta$-to-$\alpha$ relaxation transition, which is independent of the applied pressure (**Fig. 2D**). In this region, both $T_f$ and the structural parameters are strongly influenced by $T_{comp}$. This is manifested by a reduced difference in $T_f$ with respect to the reference data at 1 atm, while the glass structure begins to move toward a more compact and homogeneous configuration. When $T_{comp}$ exceeds $T_{g,P}$, the complete activation of the $\alpha$-relaxation process ensures sufficient relaxation of the sample before decompression, resulting in a glass with enhanced stability, as signaled by the lower value of $T_f$ with respect to the reference data at 1 atm. In this region, $T_f$ is also independent on $T_{comp}$ since the system reaches the liquid state and all glasses are subsequently vitrified at the same temperature. The enhanced stability structurally corresponds to permanent densification. The analysis of the $G(r)$ further reveals that glass densification is reflected in the reduction in the radial distances of different atomic coordination shells. On the short-range scale, this manifests as a significant redistribution of Cu-Zr and Zr-Zr atomic pairs, along with the shortening of Zr-Zr bonds and the consequent readjustment of the Cu-Zr bonds. As discussed by Dmwoski *et al.* (53), this structural densification represents a transition from a covalent-like bond dominated to a densely and randomly packed atomic configuration, accompanied by a reduction in structural distortion within the Zr-centered coordination shells.

Our pressure-dependent study shows also that varying pressure in the $\beta$-dominated regime results in a pressure-independent structural reorganization in the glass leading to a different state with respect to the pristine material (**Fig. 6A**). In contrast, the $\alpha$-relaxation process enhances glass stability through a pressure-dependent densification process (**Fig. 6B**), involving larger-scale structural rearrangements under high pressure. The crossover between these two different structural mechanisms is the reason of the rapid lowering of $T_f$ for compressions close to the supercooled liquid and the transition toward a homogeneous and ordered glass (**Fig. 4D** and **4F**). Additionally, by comparing with the *in-situ* XRD data (**Fig. 6C**), we also evaluated the compressibility and the potential for permanent densification of this system within the studied pressure range.

These results elucidate the role of pressure and temperature on the structure and stability in a prototypical MG, and reveal a fundamental distinction in the response of MGs to heat treatments in the presence of hydrostatic compressions. At ambient pressure, the microscopic structural differences corresponding to the $\beta$- and $\alpha$-relaxation processes are difficult to distinguish (25,26,73,80). In contrast, at high pressure one needs to take into account also the different mechanisms of structural reorganization induced by compression in the two relaxation regimes. This knowledge of the response of MGs to thermo-mechanical protocols can be used to design glasses with different properties. Applying our approach to systems exhibiting polyamorphic transitions (51,52,81) can also provide fundamental insights into the role played by the different relaxation processes during pressure-induced glass-glass or liquid-liquid transitions. This is another important implication of our work with relevant applications in the development of new amorphous materials or in their use under external stimuli.

Finally, it is important to note also the implications of our experimental approach for the study of the pressure dependence of the relaxation spectrum in glass formers. In most of MGs (including Vit4) $\beta$-relaxation is manifested by an excess wing in the dynamic mechanical spectrum (9,11,22), which makes it difficult to distinguish it from $\alpha$-relaxation. In calorimetry, this localized motion is assumed to be responsible for the kinetic transformation resulting in the observed sub-$T_g$ endothermic peak (56,57,82,83), whose intensity can be amplified under fast heating conditions (61,66,67). This is particularly important for high pressure studies as our data show that the changes in the atomic environment under high pressure lowers substantially the intensity of the sub-$T_g$ endothermic peak even for data acquired at the same degree of equilibration $T/T_{g,P}$ (**Fig. 3**). The fast-heating rates employed during the calorimetric scans provide a unique opportunity to characterize this event under high-pressure and to calculate its activation energy across a wide range of heating rates. We are confident that the experimental approach used in our work can be readily applied to other glass systems, such as the Pd-based MGs exhibiting a distinct sub-$T_g$ endothermic peak (61), and also inorganic (58) and polymeric glasses (59), to uncover new insights into the relaxation spectrum under high pressure.

## MATERIALS AND METHODS

### Sample preparation and compression protocols

Element metals with a minimum purity of 99.99% were melted under a high-purity argon atmosphere to form the master alloy, following an atomic ratio of $Zr_{46.8}Ti_{8.2}Cu_{7.5}Ni_{10}Be_{27.5}$. Ribbons of the corresponding MGs with a thickness of 30±2 $\mu m$ were then produced by melt spinning the master alloy molten liquid under Ar atmosphere. The compression protocols were performed with a Belt press at the institute Néel, Grenoble, France. For this purpose, the ribbons were cut into 2×2 mm$^2$ H×V pieces and each compression was performed by inserting 10-20 pieces in a boron nitride (BN) capsule in different groups separated by small disks of BN. The capsules were then inserted in a graphite furnace and mounted in the press, ensuring hydrostatic compression. Different densification protocols were performed in order to study separately the dependence of the material properties on the densification pressure $P_{comp}$ in the (1-7) GPa pressure range at a fixed densification temperature, $T_{comp}$, and then the dependence of $T_{comp}$ in the 298-693 K range while keeping $P_{comp}$ fixed. For each sample, we first increased the pressure to $P_{comp}$ and subsequently increased the temperature to $T_{comp}$ for a 10-minute isotherm. The samples were then cooled to 298 K under pressure, and the pressure was released only after reaching ambient temperature. The heating/cooling rate and the compression/decompression rates were set to 20 K/min and 0.3 GPa/min, respectively.

### Synchrotron XRD on compressed samples

The structure of the glasses recovered after the compression was measured at ID13 and ID15a beamlines at ESRF, Grenoble, France. At ID13, the X-ray beam energy was set at 13 keV and the diffraction data were collected using the Eiger X 4M detector using a beam size of 2.5×2.5 H×V $\mu m^2$. The detector-to-sample distance was set to 81.6 mm to achieve a detectable $q$-range of 0.1-7 $Å^{-1}$, and ensuring sufficient resolution for the first diffraction peak with one point per 0.003 $Å^{-1}$. A calibration was performed using a standard $\alpha$-$Al_2O_3$ material. Samples were mounted on a 3D-printed epoxy resin platform and kept at room temperature. We used a fixed-position optical microscope to pre-focus the different samples to ensure that the distance from each sample to the detector remained consistent, with the error being the thickness of the sample. To ensure robust statistical data, we collected approximately 1800 points of the sample with a 5 $\mu$m interval between each point and an exposure time of 0.02 s per point. Subsequently, the diffraction intensities from all points were averaged, and the background noise generated by the air adjacent to the sample was subtracted. These data provided high-resolution diffraction profiles, and all comparisons of the $q_1$ and FWHM of the FSDP discussed in the article are based on this dataset. The fitting approach for the data can be found in **Fig. S8** and **S9**.

Direct space analysis was instead performed by measuring the pair distribution functions, $G(r)$, at beamline ID15A using a 68.5 keV photon energy and a 145 mm detector-to-sample distance, probing a 0.3-30 $Å^{-1}$ detectable $q$-range. $CrO_3$ was used for calibration prior to the measurements. The samples were fixed using two parallel copper clamps. The scattered intensity was recorded with a Pilatus3 X CdTe 2M detector and each sample was exposed for 1 min to obtain good statistics. Diffraction patterns were azimuthally integrated using routines from the pyFAI library, and locally implemented corrections for the outlier rejection, background, polarization of the X-rays and detector geometry, response, and transmission, to yield 1D diffraction patterns. The structure factor, $S(q)$, was extracted from the scattered intensity $I^C(q)$ as $S(q)=1+\frac{I^C(q)-|\langle f(q)\rangle|^2}{|\langle f(q)\rangle|^2}$, where $\langle f(q)\rangle=\sum_\alpha c_\alpha f_\alpha(q)$ with $f_\alpha(q)$ and $c_\alpha$ the atomic form factor and atomic concentration for chemical species $\alpha$, respectively. $G(r)$ were then obtained through a Fourier transform of the structure factor using the equation $G(r)=\frac{2}{\pi}\int_0^{+\infty} q(S(q)-1)\sin(qr)\mathrm{d}q$ by means of the PDFgetX2 software (84).

**Synchrotron high-pressure XRD**

Additional *in-situ* high-pressure XRD measurements were conducted at P10 beamline at PETRA III in Hamburg, Germany. A pre-cut sample with the dimension of 40×40×30 H×V×L $\mu m^3$ was loaded in a membrane diamond anvil cell (DAC) with helium as a pressure-transmitting medium, and the pressure was monitored from the fluorescence spectrum of a ruby sphere placed next to the sample. The X-ray energy was set to 15.0 keV and data were recorded using an EigerX 4M detector situated 1840 mm downstream from the sample. The pressurization/decompression rate was controlled by an automatic pressure driver with a rate of 0.2 bar/s. After reaching each isobaric stage, the pressure was maintained for at least 30 minutes with a stability better than 0.1 GPa and XRD data were continuously collected during this stage, with an exposure time of 0.5 s/frame.

**Calorimetry**

Thermal properties were studied using a FDSC (flash differential scanning calorimetry, Mettler Toledo, FDSC2+) apparatus with a heating rate of 100-1000 K/s. Measurements were conducted under a high-purity $N_2$ flow of 80 ml/min. The measured samples were cut into pieces of ~80×80×30 $\mu m^3$ and then transferred to the UFS chip of the calorimeter. Unlike the traditional melting-quenching-heating measurements, we used a single-shot method for thermal analysis, with each sample being used only once. For each sample, at least two different sections were cut and measured independently to ensure reproducibility (**Fig. S1**).

In order to determine the mass of the sample in FDSC, we first conduct a DSC (Mettler Toledo, DSC3) measurement by using approximately 10 mg of the sample to assess the heat flow step, $\Delta Q$, which adheres to the formula $\Delta Q = C_p * \Phi * m$, where $C_p$ represents the specific heat of the sample, $\Phi$ is the heating rate, and $m$ is the sample mass. By ascertaining the $\Delta Q$ in FDSC, we can further verify the mass of the sample with the FDSC setup based on the sample $C_p$.

Fictive temperatures, $T_{\mathrm{f}}$, were calculated using Moynihan's area-matching method (62) with the following equation: $\int_{T_{\mathrm{f}}}^{T_1 \gg T_{\mathrm{g}}} (C_{\mathrm{p,\ liquid}} - C_{\mathrm{p,\ glass}}) dT = \int_{T_2 \ll T_{\mathrm{g}}}^{T_1 \gg T_{\mathrm{g}}} (C_{\mathrm{p}} - C_{\mathrm{p,\ glass}}) dT$, where $C_{\mathrm{p,\ liquid}}$ and $C_{\mathrm{p,\ glass}}$ represent the specific heat capacities of the liquid and glass, respectively.

The endothermic peaks with different characteristic temperatures, $T_{\mathrm{p}}$ (**Fig. S2** shows the definition of $T_{\mathrm{p}}$), of the FDSC curve indicate the kinetic transformation associated to relaxation processes. The Kissinger equation (85) of the different $T_{\mathrm{p}}$ and the $\Phi$ were used to estimate the activation energies, $E_{\mathrm{a}}$, of these processes: $\ln\left(\frac{\Phi}{T_{\mathrm{p}}^2}\right) = -\frac{E_a}{RT_{\mathrm{p}}} + a$, where $R$ is the gas constant and $a$ is a constant.

**Acknowledgements:** We gratefully acknowledge the ESRF (Grenoble, France) for providing beamtime, including experiments conducted during in-house beamtime (ihsc1815) at ID13 and the beamtime under the proposal HC4986 at ID15A. We also thank DESY (Hamburg, Germany), a member of the Helmholtz Association HGF, for providing the beamtime under the proposal 11016133EC beamline P10 at PETRA III. **Funding:** This work received funding from the European Research Council (ERC) under the European Union's Horizon 2020 research and innovation program (Grant Agreement No. 948780). **Author contributions:** JS, AC, CG, and ML conducted the densification experiments. FY and EP provided the samples. JS, AC, AR, BR, GV, and MdM performed the PDF experiment, and GV and JS analyzed the data. JS and JL carried out the XRD experiment and analyzed the data. JS conducted the FDSC experiments and analyzed the results with the assistance of DC, JG, and GM. JS, AC, AR, and BR conducted the *in-situ* high-pressure XRD experiments with support from GG and FW. JS and BR wrote the manuscript with contributions from all authors. JS and AC contributed equally to this work. BR conceived the study and led the investigation. **Competing interests:** All authors declare that they have no competing interests. **Data and materials availability:** All data needed to evaluate the conclusions in the paper are present in the paper and/or the Supplementary Materials.

# Figure legends:

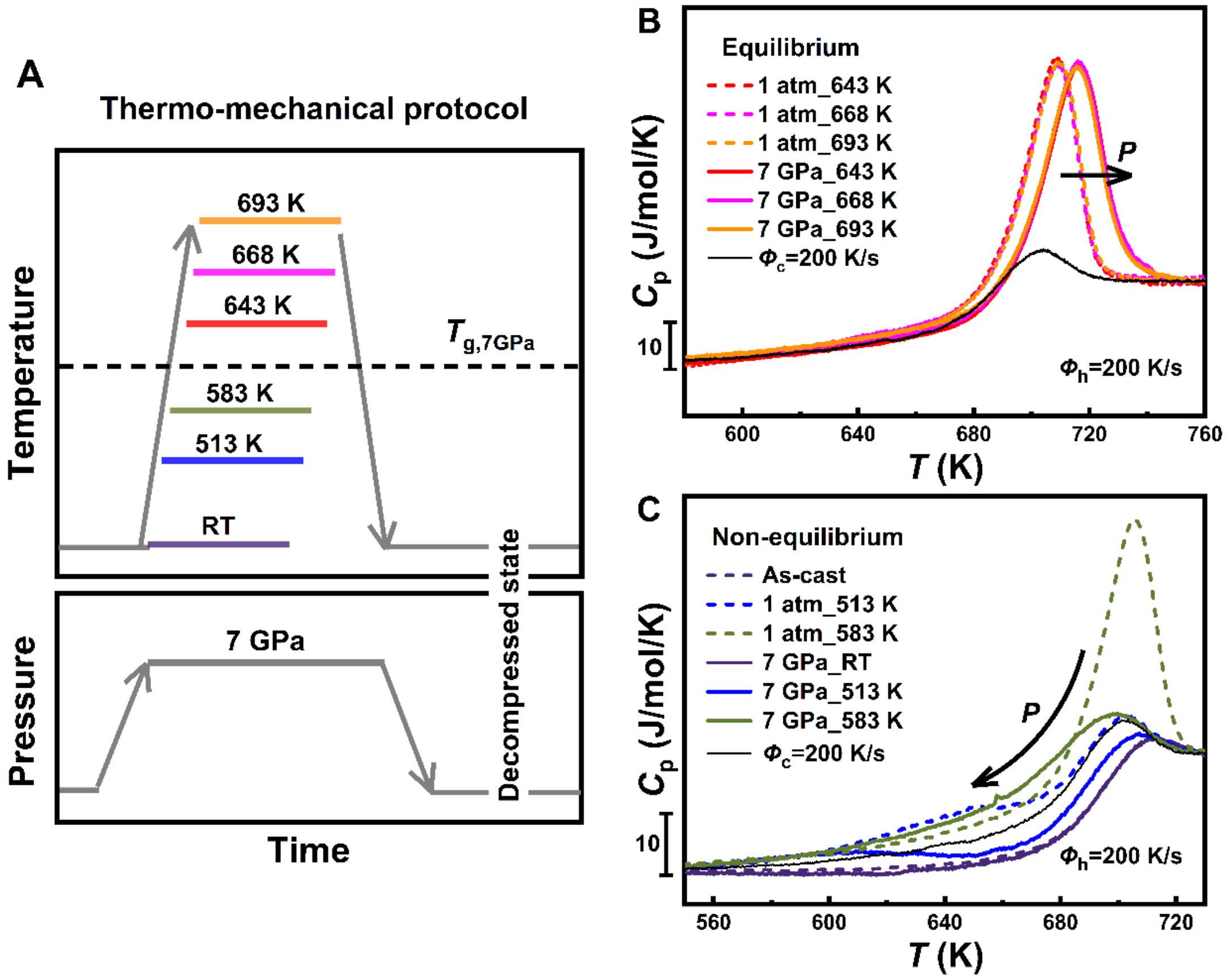

**Figure 1. The thermodynamic state of the compressed glasses. (A)** Sketch of the thermo-mechanical protocol. First pressure is increased at $P_{comp}$=7 GPa, and then for each sample the temperature is raised to a fixed value of $T_{comp.}$ ranging from 298 K to 693 K for 10 min. When $T_{comp} > T_{g,P}$, the sample under compression is in thermodynamic equilibrium; when $T_{comp} < T_{g,P}$, it is in the non-equilibrium glass state. The temperature is then cooled back to the ambient value with a rate of 20 K/min prior to the release of the pressure. The resulting glasses are then recovered at ambient condition for calorimetry measurements. **(B)** and **(C)** Selection of calorimetry heating curves measured with FDSC at 200 K/s for samples previously compressed at 7 GPa and $T_{comp} \geq 643$ K **(B)** and at $T_{comp} \leq 583$ K **(C)**. Reference data of samples pre-annealed at annealing temperatures $T_a$=$T_{comp}$ and ambient pressure (dashed lines) are reported as well for comparison, together with data acquired in a glass cooled from the supercooled liquid state with a rate $\Phi_c$ =200 K/s equal to the heating rate used immediately after for the measurements (black line).

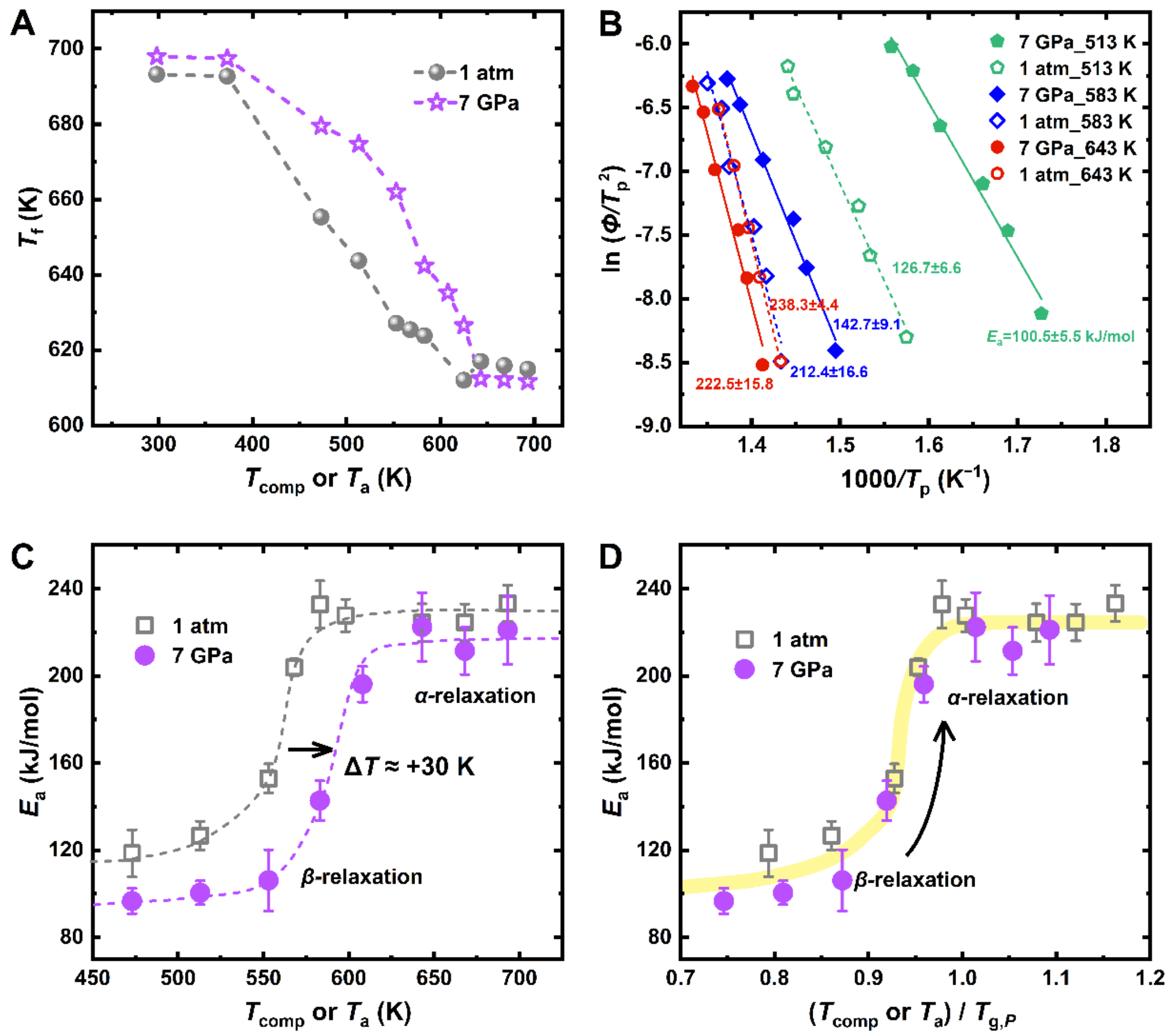

**Figure 2. Pressure dependence of the thermal stability and relaxation spectrum. (A)** Fictive temperature, $T_f$ of different annealed and compressed glasses as a function of $T_a$ or $T_{comp}$. **(B)** Kissinger plot illustrating the characteristic temperature, $T_p$, of the endothermic peak (see **Fig. S2** for definition) in the FDSC curves as a function of the heating rate $\Phi_h$ for different annealed and compressed glasses. **(C)** Activation energy, $E_a$ of different annealed and compressed glasses. The data are reported as a function of $T_{comp}$ or $T_a$. Only data displaying a clear endothermic peak have been used for the analysis of the activation energies. **(D)** $E_a$ of different annealed and compressed glasses as a function of $T_{comp}$ and $T_a$ normalized by the corresponding $T_{g,P}$. The arrow indicates the transition from $\beta$- to $\alpha$-relaxation.

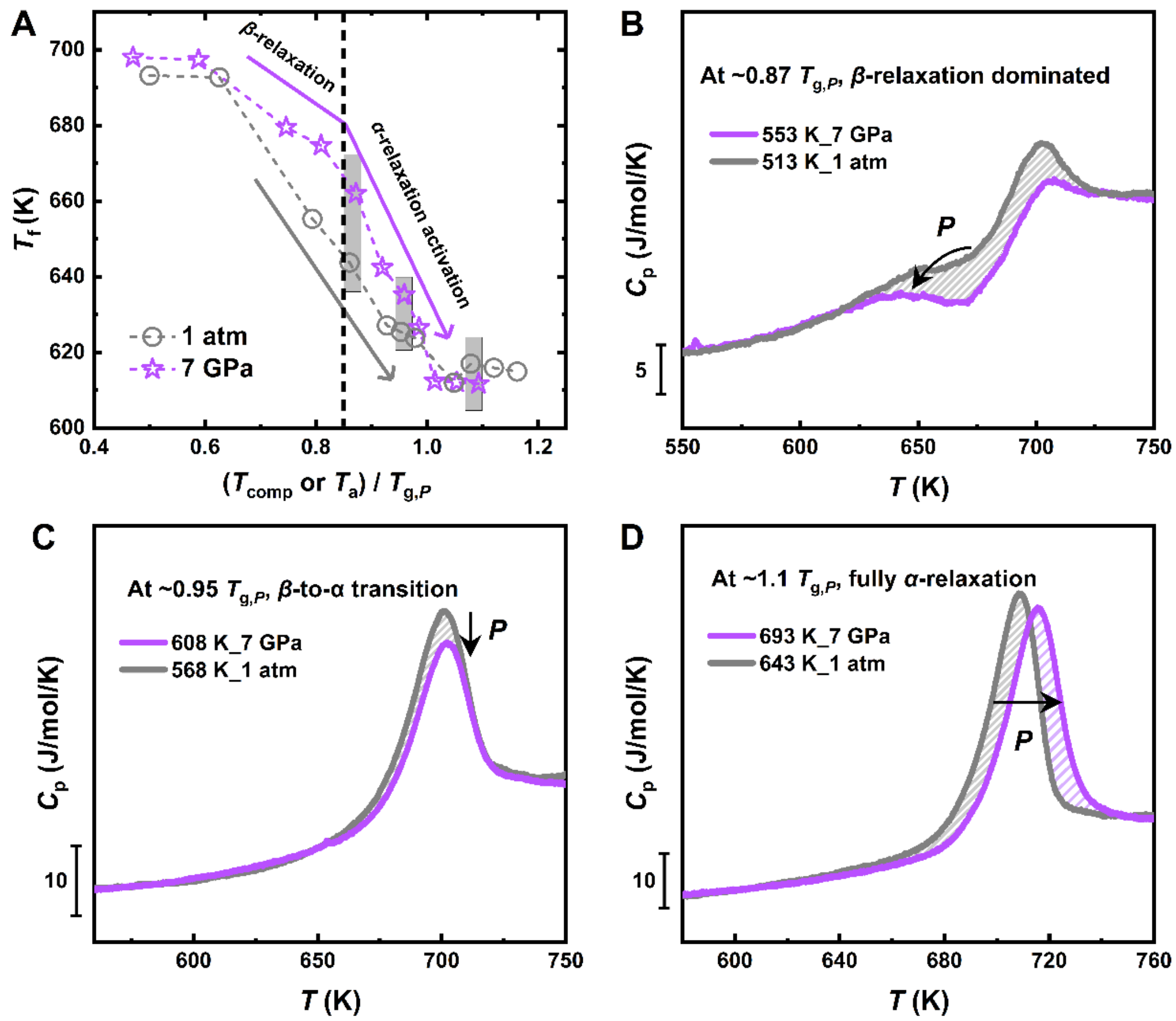

**Figure 3. Pressure response of $T_f$ at the same degree of equilibration. (A)** Fictive temperature, $T_f$ of different annealed and compressed glasses as a function of $T_{comp.}$ and $T_a$ normalized by the corresponding $T_{g,P}$. The arrows indicate the evolution trend of $T_f$ within the dynamical ranges corresponding to the $\beta$- and $\alpha$-relaxation, while the black line corresponds to the onset transition between these processes, as determined from the activation energy analysis. **(B, C, D)** Comparison of FDSC curves for compressed and 1 atm-annealed samples at temperatures corresponding to the same degree of equilibration of ~0.87 $T_{g,P}$, ~0.95 $T_{g,P}$, and ~1.1 $T_{g,P}$, respectively, which correspond to the $T_f$ data marked by the boxes in panel (A).

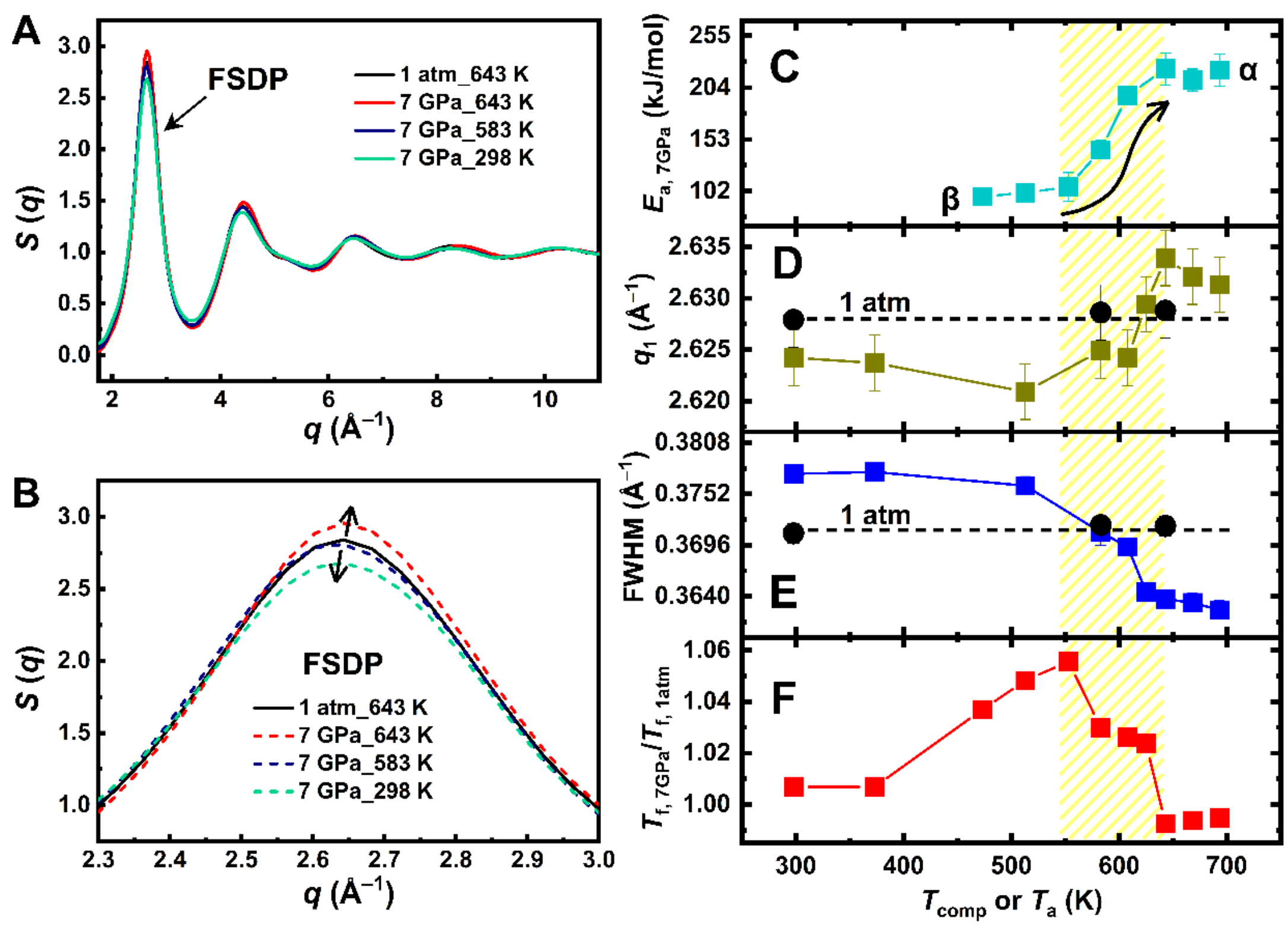

**Figure 4. Structural properties of annealed and compressed glasses. (A)** Static structure profiles, $S(q)$, measured at ambient temperature with synchrotron XRD in glasses previously compressed at 7 GPa and different $T_{comp}$. A reference sample annealed at 643K and 1 atm is also reported for comparison. **(B)** Zoom of the top part of the first sharp diffraction peak (FSDP). **(C)-(F)** Evolution of calorimetric and structural parameters of glasses densified at 7 GPa and at various $T_{comp}$. Data represent: $E_a$ **(C)**, maximum position of $q_1$ of the FSDP **(D)**, FWHM of the FSDP **(E)**, and the $T_f$ normalized to the $T_g$ **(F)**. Reference data for annealed glasses at 1 atm are also included (black circles) for both $q_1$ and the FWHM. The error in $q_1$ is determined by the maximum deviation (sample thickness) in the sample-to-detector distance for each measurement (see Methods section), while the error in FWHM arises from the fitting uncertainties. The shaded area highlights the temperature transition region from $\beta$-relaxation to $\alpha$-relaxation determined by the calorimetric measurements.

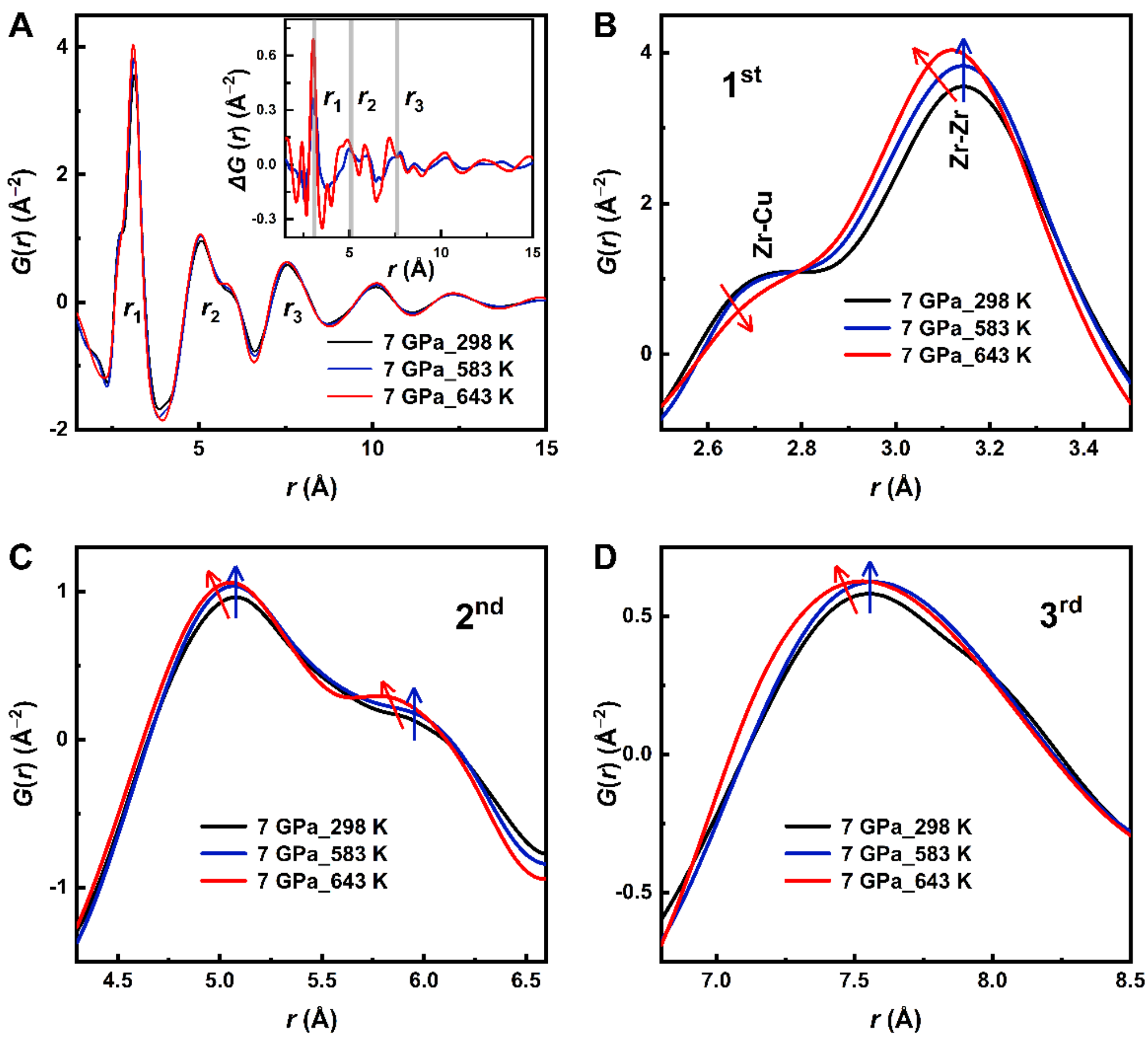

**Figure 5**. **Real space analysis of the structure after compression at different $T_{comp}$. (A)** Pair distribution function, $G(r)$, of Vit 4 compressed at 7 GPa and $T_{comp}$ = 298 K (cold compression), 583 K (onset crossover $\beta$-to-$\alpha$ relaxation regimes), and 643 K ($\alpha$-relaxation regime). The insets show the difference between the $G(r)$ of the samples compressed at 583 K and 643 K, respectively, and the $G(r)$ of the cold-compressed sample. The vertical lines indicate the peak positions of the first three coordination peaks of the cold-compressed sample. **(B-D)** Close-up views of the $1^{st}$ to $3^{rd}$ atomic coordination peaks of $G(r)$. The blue and red arrows indicate the changing direction of $G(r)$ for the samples compressed at 583 K and 643 K, respectively, relative to the cold-compressed sample. The sub-peak and main peak in (B) are primarily corresponding to Zr-Zr and Cu-Zr atomic pairs [74], respectively.

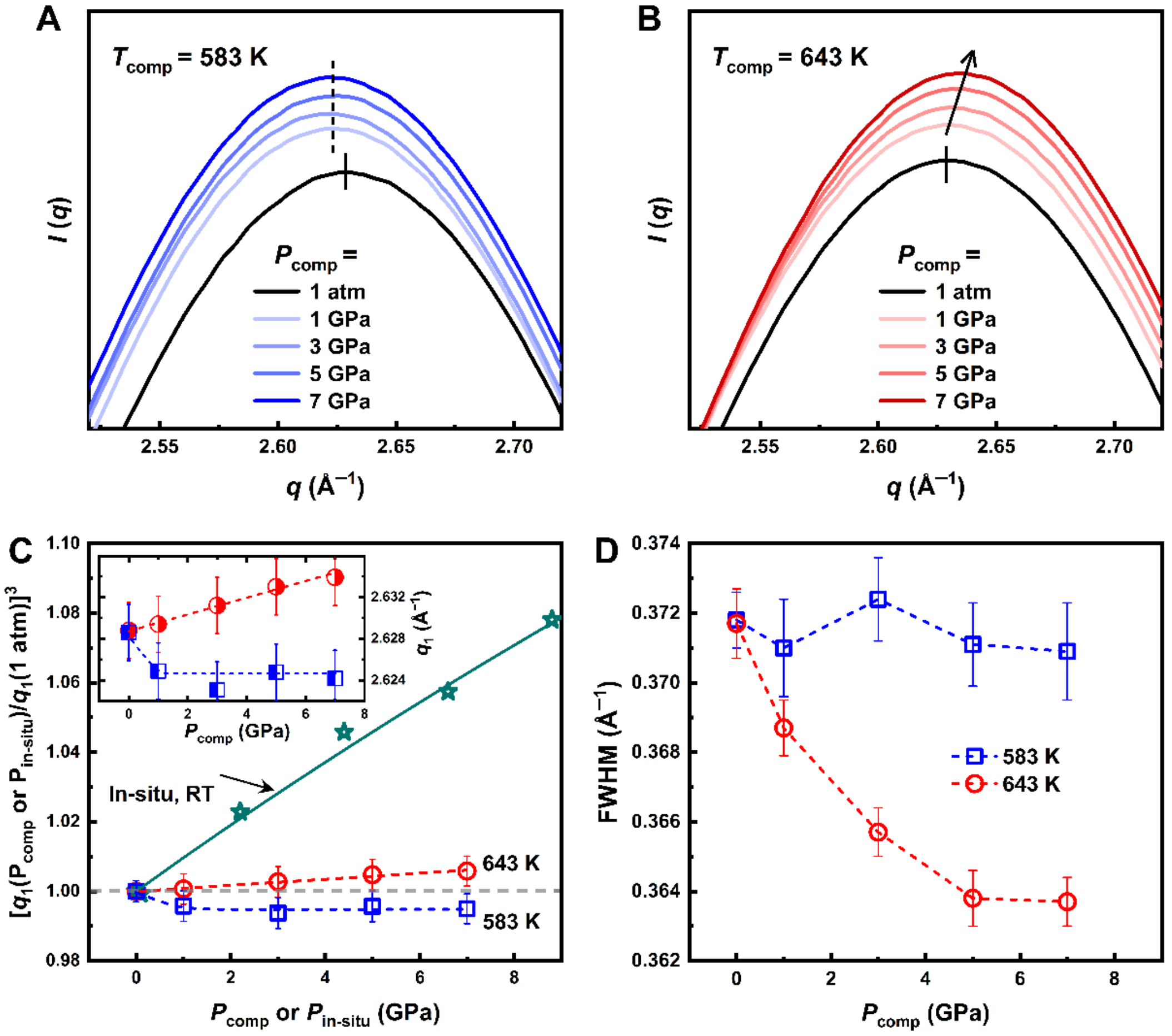

**Figure 6. Pressure dependence of the FSDP.** Intensity profile of glass samples compressed at different pressures and then measured at ambient temperature using synchrotron XRD. Measurements were performed for $T_{comp}$ below $T_{g,1atm}$=596 K **(A)**, and above $T_{g,7GPa}$=634 K **(B)**. To differentiate the curves, a vertical offset is applied. Lines and arrows indicate the evolution of peak positions with increasing $P_{comp}$. **(C)** Pressure dependence of the FSDP $q_1^3$ for a sample kept *in-situ* at different pressures values $P_{in-situ}$, and for the glasses of panel (A) and (B) previously compressed at different $P_{comp}$ and $T_{comp}$ and measured here at 1 atm. All data are rescaled by the reference value at 1 atm. The difference between the *in-situ* and *ex-situ* densifications provides an indication of the volume recovery after the decompression. The green line is the fit to the third-order BM-EOS. The inset reports $q_1$ value for different compressed glasses of panel (A) and (B). **(D)** Evolution of the FWHM of the XRD data of panel (A) and (B).